\documentstyle[12pt]{article}
\def\beq{\begin{equation}}
\def\eeq{\end{equation}}

\begin{document}
\title{Gravitational waves interacting with a spinning charged
particle in the presence of a uniform magnetic field}
\author{D. B. Papadopoulos\thanks{E-mail address:papadop@astro.auth.gr}\\
Department of Physics, \\
\small Section of Astrophysics, Astronomy and Mechanics, \\
\small Aristotelian University of Thessaloniki, \\
\small 54124 Thessaloniki, GREECE}

\maketitle

\begin{abstract}The equations which determine the response of a spinning
charged particle moving in a uniform magnetic field to an incident
gravitational wave are derived in the linearized approximation to
general relativity. We verify that 1) the components of the
4-momentum, 4-velocity and the components of the spinning tensor,
both electric and magnetic moments, exhibit resonances and 2) the
co-existence of the uniform magnetic field and the GW are
responsible for the resonances appearing in our equations. In the
absence of the GW, the magnetic field and the components of the
spin tensor decouple and the magnetic resonances disappear.
\end{abstract}

\maketitle

\section{Introduction}

In recent years, there exists an increasing interest in problems
related to the motion of spinning particles in the presence of
external fields (e.g. gravitational and electromagnetic fields).
Thus, some interesting results related to this topic have been
discussed by Mohseni et. al.[1,2,3]. Actually, they have
discussed the motion of a classical spinning particle in the
field of a weak gravitational wave (GW) and they found that the
characteristic dimensions of the particle's orbits are sensitive
to the ratio of the spin to the mass of the particle.

The problem of spinning particle(s), in the frame of special
relativity, has been discussed by Frenkel[4] and later on,
Bargmann et.al.[5] extended Frenkel's work for a spinning
particle in the presence of an electromagnetic field. Karl Yee et
al.[6] have obtained the equations of motion for the position and
spin of a classical particle coupled to an external
electromagnetic and gravitational potential from an action
principle.

The discussion of a spinning particle in the frame of general
relativity goes back to Mathisson [7] and Papapetrou [8] (For a
recent discussion on Mathisson's spinning electron equations see
also Horvathy [9]). Neglecting self-gravitation and back reaction
they have endowed a particle with spin by considering a rotating
mass energy distribution in the limit of vanishing volume but
with the angular momentum remaining finite. Later on the theory
was clarified by Dixon [10] and developed by Ehlers and Rudolph
[11].

Cho[12], using an energy-momentum tensor for spinning particles
due to Dixon and Bailey-Israel[13], developed the post-Newtonian
approximation for n spinning particles in a self-contained manner
following closely the procedure presented in the well known text
book by Weinberg [14].

The motion of a classical charged massive spinning particle in
the frame of general relativity in the presence of an external
electromagnetic and gravitational field was extensively
investigated by Dixon [15,16] and Souriau [17,18,19]. The
dynamics of spinning test particles which are close to massive
compact objects was investigated by several authors; this kind of
investigation includes, generation of gravitational waves as
spinning particles fall into black holes [20,21,22], chaotic
behavior of spinning particles in certain space times [23,24,25]
and numerical computations for orbital motions in Kerr background
[26]. Interesting results regarding the interaction of spinning
test matter with gravitational and electromagnetic waves have
been obtained by studying the classical motion of the spinning
test particles in plane gravitational and electromagnetic field
solution to the Einstein-Maxwell equations [27]. Plane-fronted
gravitational waves (pp-waves) have received some attention
recently because of their high symmetry and the fact that
gravitational plane waves are a subclass of pp-waves [28,2]. They
are assumed to describe the gravitational field at great
distances from the radiating source and they can be purely
gravitational, purely electromagnetic or both, depending on the
source. In such backgrounds scattering effects of spinning
particles have been discussed from a different point of view by
several authors [27,28,29].

Using Grassmann variables Barducci et. al.[30], Ravndal [31] and
Rietdijk and van Holten[32] have obtained several interesting
results related to the spinning particles. Actually, in a paper
by van Holten [33], the constrain $p^aS_{ab}=0$, is satisfied by
expressing the spin tensor $S_{ab}$ as a product of two Grassmann
variables. In this case, the equations of motions are derived
using a Dirac-Poisson brackets formalism.

In this paper, we consider a charged massive spinning particle,
in the Dixon-Souriau(DS) model[21], which includes spin-gravity
terms and spin-electromagnetism terms as well (for an uncharged
particle the model reduces to the well known Papapetrou one).
Following the strategy of Ref.(1), in the so called DS equations
of motion and neglecting the spin-electromagnetism interaction
terms [21,35], we determine the response of a spinning charged
particle in a uniform magnetic field to an incident GW and verify
a strong coupling of the external fields (magnetic and
gravitational fields) to the spin, producing resonaces. Those
resonances may cause drastic enhancement in motion which may have
interesting astrophysical applications.

The paper is organized as follows: In Section 2, we present the
so called DS equation of motion and describe the components of the
GW which disturbs the Minkowski space time with signature
$(-1,1,1,1)$ . In Section 3, we consider only the case where
$\lambda=0$. In other words, in the DS equations we neglect the
spin-electromagnetic interaction terms and in the frame of
linearized theory of gravity we obtain, for the spinning charged
particle, solutions for the components of the 4-momentum, its
4-velocity and its space components $x^{\mu}$ in a coordinate
system $(t,x,y,z)$. In Section 4, we discuss the obtained results.

\newpage

\section {The DS Equations of Motion}

The equations of motion of a spinning test particle originally
derived from Papapetrou [8] and later on reformulated by Dixon
[10,15,16].

Souriau [18,19] derived the so called DS equations of motion of a
spinning test particle with charge e in a given gravitational and
electromagnetic background. These equations are [21]:
\begin{equation}\label{vel1}
\frac{d x^{\mu}}{d \tau}=\upsilon^{\mu}
\end{equation}
\begin{equation}\label{momen1}
 \frac{d p^{\mu}}{d \tau}=-\Gamma_{\lambda
\nu}^{\mu}\upsilon^{\lambda}
p^{\nu}-\frac{1}{2}R^{\mu}_{\nu\lambda\rho}
S^{\lambda\rho}\upsilon^{\nu}+eF_{\beta}^{\mu}\upsilon^{\beta}-\frac{\lambda}{2}S^{\kappa\rho}\nabla^{\mu}
F_{\kappa\rho} \end{equation}

\begin{equation}\label{spin1}
\frac{dS^{\mu\nu}}{d\tau}=-\Gamma_{\lambda\rho}^{\mu}\upsilon^{\lambda}S^{\rho\nu}-\Gamma_{\lambda\rho}^{\nu}\upsilon^{\lambda}S^{\mu\rho}+(p^{\mu}\upsilon^{\nu}-p^{\nu}\upsilon^{\mu})+\lambda(S^{\mu\kappa}
F_{\kappa}^{\nu}-S^{\nu\kappa}F_{\kappa}^{\mu})
\end{equation}
where Greek indices take values 0,1,2,3, Latin 1,2,3, $\tau$ is
an affine parameter across a word line L which is chosen as the
proper time of the charged particle, $\upsilon^{\mu}$ is the
4-velocity of the charged particle across the world line L,
$p^{\mu}=\int T^{0\mu}dV$ are the components of the 4-momentum of
the spinning charged particle, $F^{\mu\nu}$ is the
electromagnetic tensor, $\lambda$ is an electromagnetic coupling
scalar and $S^{\mu\nu}$ is the spin tensor. Unlike special
relativity, $p^{\mu}$ and $\upsilon^{\mu}$ are not generally
proportional to each other. But it is well known that
Eqs.(\ref{momen1}) and (\ref{spin1}) themselves, do not constitute
an independent set of equations since  they are less than the
unknown quantities (3 components of the spin tensor are not
determined). Therefore, several supplementary conditions have
been used in the literature to remedy this problem [36]. Here we
will adopt Dixon's condition [15] e.g.
$p_{\mu}S^{\mu\nu}=0$(center of mass condition).

To find  the trajectory of the spinning charged particle, we need
to know its 4-velocity. But there are no equations of motion for
this purpose. However, we may obtain indirectly a relation between
$\upsilon^{\nu}$ and $p^{\nu}$ from the following equation [1]:

\begin{eqnarray}\label{vel2}\upsilon^{\nu}&=&N\{
u^{\nu}-\frac{1}{2m^2\Delta}R_{\mu \beta
\lambda\rho}S^{\lambda\rho}S^{\mu\nu}u^{\beta}+\frac{e}{m^2\Delta}S^{\mu\nu}F_{\mu
\beta}u^{\beta}\nonumber\\
&+&\lambda[\frac{1}{2m^3\Delta}R_{\mu
\beta\lambda\rho}S^{\lambda\rho}S^{\mu\nu}S^{\beta
\kappa}(F_{\kappa}^{\sigma}
u_{\sigma})-\frac{1}{2m^2\Delta}S^{\mu\nu}S^{\kappa\rho}\nabla_{\mu}F_{\kappa\rho}-\frac{1}{m}S^{\nu\kappa}(F_{\kappa}^{\sigma}
u_{\sigma})]\nonumber\\
&-&\frac{\lambda e}{m^3\Delta}S^{\mu\nu}F_{\mu \beta}S^{\beta
\kappa}(F_{\kappa}^{\sigma} u_{\sigma})\}\end{eqnarray} where
$p^{\mu}=mu^{\mu}$, $p_{\mu}p^{\mu}=-m^2$,  $m$ is the mass of the
particle and $G=c=1$
\begin{equation}\label{delta}
\Delta=1+\frac{1}{4m^2}R_{t \mu\lambda\rho}S^{\lambda\rho}S^{t
\mu}+\frac{1}{2m^2}eF_{t \mu}S^{t \mu} \end{equation}

\begin{eqnarray}\label{Nu}
N&=&\{1-\Lambda_{\mu}^{(1)}\Lambda_{\alpha}^{(1)}S^{\mu\nu}S_{\nu}^{\alpha}-2\lambda[\Lambda_{\mu}^{(2)}\Lambda_{\alpha}^{(1)}S^{\mu\nu}S_{\nu}^{\alpha}-(F_{\kappa}^{\sigma}
u_{\sigma})\Lambda_{\alpha}^{(1)}S^{\kappa\nu}S_{\nu}^{\alpha}]\nonumber\\
&-&\lambda^2[\Lambda_{\mu}^{(2)}\Lambda_{\alpha}^{(2)}S^{\mu\nu}S_{\nu}^{\alpha}-2(F_{\kappa}^{\sigma}
u_{\sigma})\Lambda_{\alpha}^{(2)}S^{\kappa\nu}S_{\nu}^{\alpha}+(F_{\kappa}^{\sigma}
u_{\sigma})(F_{\beta}^{\alpha} u_{\alpha})S^{\kappa\nu}S_{\nu}^{\beta}]\nonumber\\
&-&2\lambda e
\Lambda_{\mu}^{(3)}\Lambda_{\alpha}^{(3)}S^{\mu\nu}S_{\nu}^{\alpha}-2e\lambda^2[-\Lambda_{\mu}^{(3)}\Lambda_{\alpha}^{(2)}S^{\mu\nu}S_{\nu}^{\alpha}+(F_{\kappa}^{\sigma}
u_{\sigma})\Lambda_{\mu}^{(3)}S^{\mu\nu}S_{\nu}^{\kappa}]\nonumber\\
&+&\lambda^2 e^3
\Lambda_{\mu}^{(3)}\Lambda_{\alpha}^{(3)}S^{\mu\nu}S_{\nu}^{\alpha}
\}^{-1/2}\end{eqnarray} and
\begin{equation}\label{Lam1}
\Lambda_{x}^{(1)}=\frac{1}{2m^2\Delta}R_{x\sigma
\lambda\rho}S^{\lambda\rho}u^{\sigma}+\frac{e}{m^2\Delta}F_{x\sigma}u^{\sigma}
\end{equation}

\begin{equation}\label{Lam2} \Lambda_{x}^{(2)}=-\frac{1}{2m^3\Delta}R_{x\sigma
\lambda\rho}S^{\lambda\rho}S^{\sigma \kappa}(F_{\kappa}^{\beta}
u_{\beta}) +\frac{1}{m^2\Delta} S^{\kappa\rho} \nabla_{x}
F_{\kappa\rho}\end{equation}

\begin{equation}\label{Lam3} \Lambda_{x}^{(3)}=\frac{1}{m^3\Delta}F_{x\sigma} S^{\sigma \kappa}
(F_{\kappa}^{\beta} u_{\beta})\end{equation} where $x$ stands for
$x=\mu,\alpha$. Upon the consideration of the
assumption($\lambda=0$) (Pomeranski\v{i} et al. 2000 and
references therein), we neglect particular terms in
Eqs.(\ref{momen1}),(\ref{spin1}) and a simplified covariant model
is obtained:

\begin{equation}\label{momen2}
 \frac{d p^{\mu}}{d \tau}=-\Gamma_{\lambda
\nu}^{\mu}\upsilon^{\lambda}
p^{\nu}-\frac{1}{2}R^{\mu}_{\nu\lambda\rho}
S^{\lambda\rho}\upsilon^{\nu}+eF_{\beta}^{\mu}\upsilon^{\beta}
\end{equation}

\begin{equation}\label{spin2}
\frac{dS^{\mu\nu}}{d\tau}=-\Gamma_{\lambda\rho}^{\mu}\upsilon^{\lambda}S^{\rho\nu}-\Gamma_{\lambda\rho}^{\nu}\upsilon^{\lambda}S^{\mu\rho}+(p^{\mu}\upsilon^{\nu}-p^{\nu}\upsilon^{\mu})
\end{equation}

\begin{eqnarray}\label{vel3}\upsilon^{\nu}&=&N\{
u^{\nu}-\frac{1}{2m^2\Delta}R_{\mu \beta
\lambda\rho}S^{\lambda\rho}S^{\mu\nu}u^{\beta}+\frac{e}{m^2\Delta}S^{\mu\nu}F_{\mu
\rho}u^{\rho}\}
\end{eqnarray}
with
\begin{equation}\label{Nu2}
N=[1-\frac{N_a}{4m^4\Delta^2}]^{-1/2}
\end{equation}
and
\begin{eqnarray}\label{Nua}
N_a&=&(R_{\mu\sigma\lambda\rho}S^{\lambda\rho}S^{\mu\nu}u^{\sigma})[R_{\mu\sigma\lambda\rho}S^{\lambda\rho}(\eta_{\nu\beta}+h_{\nu\beta})S^{\mu\beta}u^{\sigma}]\nonumber\\
&-&2e[R_{\mu\sigma\lambda\rho}S^{\lambda\rho}(\eta_{\nu\beta}+h_{\nu\beta})S^{\mu\beta}u^{\sigma}](F_{\kappa\gamma}S^{\kappa\nu}u^{\gamma})\nonumber\\
&-&2e(R_{\mu\sigma\lambda\rho}S^{\lambda\rho}S^{\mu\nu}u^{\sigma})[F_{\kappa\sigma}(\eta_{\nu\beta}+h_{\nu\beta})S^{\kappa\beta}u^{\sigma}]\nonumber\\
&+&4e^2[F_{\kappa\sigma}(\eta_{\nu\beta}+h_{\nu\beta})S^{\kappa\beta}u^{\sigma}](F_{\kappa\sigma}S^{\kappa\nu}u^{\sigma})
\end{eqnarray}
In the next section, we discuss the above equations in the
linearized theory of gravity.

\section{The DS-equations of motion in the linearized theory of gravity }

To understand the Eqs.(\ref{momen2}-\ref{Nua}) in the linearized
approximation to general relativity, we decompose the metric in
the fashion
\begin{equation}\label{metric1}
 g_{\mu\nu}=\eta_{\mu\nu}+h_{\mu\nu} \end{equation}
By imposing the condition
\begin{equation}\label{metric2}
(h_{\mu}^{\nu}-\delta_{\mu}^{\nu}
h_{\rho}^{\rho})_{;\nu}=0\end{equation} we reduce the vacuum
field equations to homogeneous wave equations for all components
of $h_{\nu}^{\mu}$. A coordinate transformation can now be
effected to reduce the trace $h_{\nu}^{\nu}$ and the mixed
components $h_{0\alpha}$, to zero. The gravitational field is
then described by a symmetric traceless, divergenceless tensor
with two independent space components which, for simplicity, we
call $h_1=h_{+}$ and $h_2=h_{\times}$. Thus, the square of the
line element is
\begin{eqnarray}\label{metric3}
ds^2&=&(\eta_{\mu\nu}+h_{\mu\nu})dx^{\mu}
dx^{\nu}\nonumber\\
&=&-(dx^0)^2+(1+h_1)dx^2+(1-h_1)dy^2+dz^2+2h_2dxdy\end{eqnarray}
where $\vert h_{1},h_2\vert \ll 1$. We consider a plane GW which
is characterized by the wave 3-vector

\begin{equation}\label{wavenum}
k_g^{i}=\omega_g(0,0,1)\end{equation} and the two possible states
of polarization given by
\begin{equation}\label{pert1}
h_{1}=h_{10}e^{i(k_gz-\omega_g
t)},~~~h_{2}=h_{20}e^{i(k_gz-\omega_g t)}  \end{equation} where
$h_{10},h_{20}$ are the amplitudes of the two components of the
GW.

We choose the electromagnetic field to be

\begin{eqnarray}\label{magnet} F_{ij}=\left (
\begin{array}{cccc}
0 & 0 & 0 & 0 \\
0 & 0 &-H_3 & 0 \\
0 & H_3 & 0 & 0\\
0 & 0 & 0 & 0
\end{array}
\right ) \end{eqnarray} where the background magnetic field is
constant e.g $H^a=(0,0,H^3)=const.$(from now on $H^3=H$). In this
scenario, the metric is still a solution to the Einstein equations
in vacuum, because we assume that the energy density of the
magnetic field, is approximately zero(i.e. no effect of the
magnetic field on either the $\eta_{\mu\nu}$ or $h_{\mu\nu}$).

We intend to discuss the electrodynamics of a spinning point-like
charged particle with mass m, with an intrinsic angular
momentum,  in the presence of a uniform magnetic field across the
z-axis, initially at rest with respect to the coordinate system
in which the metric (\ref{metric3}) is expressed. To achieve this
task, a relation between the invariant proper time $\tau$ and the
coordinate time $t$ is needed. In the absence of external forces,
this relation may be found from the expression
\begin{equation}\label{prop1}
ds^2=(\eta_{\mu\nu}+h_{\mu\nu})dx^{\mu}dx^{\nu}=-d\tau^2\end{equation}
because in Einstein's theory of gravity, the word lines of
classical point particles in curved space times are time-like
(see Ref.[33,34]. In the case we have external forces, we have to
use the expression
\begin{equation}\label{prop2}
d\tau=dt[\frac{1-\upsilon^2}{1-eS^{\mu\nu}F_{\mu\nu}/m^2}]^{1/2}\end{equation}
where  $\upsilon^2$ is the total space velocity of the spinning
particle. Looking at the Eq.(22) we would like to make the
following comments: a)The physical meaning of the Eq.(22) is that
relativistic time dilation occurs for a spinning charged particle
with non-zero magnetic moments in an external magnetic field. b)
The structure of Eq.(22) results from the fact that
$\upsilon^{\mu}$ and $u^{\mu}$ differ from each other, where
$\tau$ is connected with $\upsilon^{\mu}$ and where $u^{\mu}$ is
normalized [33,34].

To make some further progress with the
Eqs.(\ref{momen2}-\ref{Nua}), we decompose the particle's
components of the 4-velocity, 4-momentum and spin tensor as
follows:

\begin{eqnarray}\label{pert2}
\upsilon^{\mu}\simeq
\upsilon_{0}^{\mu}+\upsilon_{1}^{\mu},~~p^{\mu}\simeq
p_{0}^{\mu}+p_{1}^{\mu},~~S^{\mu\nu}\simeq
S_{0}^{\mu\nu}+S_{1}^{\mu\nu}\end{eqnarray} with
$\upsilon_{1}^{\mu}$, $p_{1}^{\mu}$ and $S_{1}^{\mu\nu}$ being of
the same order as $h_{\mu\nu}$.

Thus, from Eqs.(\ref{momen2}-\ref{Nua}) and with the aid of
Eqs.(\ref{prop2}-\ref{pert2}) we obtain the following equations:

{\bf 3.a) Zero Order Equations}

\begin{equation}\label{eq01}
 \frac{d u_0^{\mu}}{d t}=\frac{e}{m}\eta^{\mu\nu}F_{\nu\beta}\upsilon_0^{\beta} \end{equation}

\begin{equation}\label{eq02}
\frac{dS_0^{\mu\nu}}{d
t}=m(u_0^{\mu}\upsilon_0^{\nu}-u_0^{\nu}\upsilon_0^{\mu})
\end{equation}

\begin{eqnarray}\label{eq03}\upsilon_0^{\nu}&=&[\frac{1}{1-eS^{\mu\nu}F_{\mu\nu}/m^2}]^{1/2}[1-\frac{e^2N_0}{m^4\Delta_0}]^{-1/2}\{
u_0^{\nu}+\nonumber\\
&+&\frac{e}{m^2\Delta_0}S_0^{\mu\nu}F_{\mu
\sigma}u_0^{\sigma}\}\end{eqnarray} where
\begin{equation}\label{eq04}
\Delta_0=1+\frac{1}{2m^2}eF_{\sigma \mu}S_0^{\sigma \mu}
\end{equation}

\begin{equation}\label{eq05}
N_0=[F_{\kappa\sigma}S_0^{\kappa\nu}u_0^{\sigma}][F_{\kappa\sigma}\eta_{\nu\beta}S_0^{\kappa\beta}u_0^{\sigma}]
\end{equation}
We assume that the spinning particle initially is at rest,
$u_0^{\mu}=(1,0,0,0),\upsilon_0^{\mu}=(1,0,0,0)$; then the
condition $p_{\mu (0)}S_0^{\mu\nu}=0$, implies that the zero
order electric moments of the spin-tensor vanish e.g
$S_0^{0\nu}=0$. After some straightforward calculations and with
the aid of Eqs.(\ref{prop2}),(\ref{eq04}) and (\ref{eq05}) we
find; $\Delta_0=1-\frac{eH}{m}(\frac{S_0^{12}}{m})$ and, for the
system (\ref{eq02}), the following zero-order equations are
derived:
\begin{equation}\label{eq06}
S_0^{12}=constant,S_0^{13}=constant,S_0^{23}=constant
\end{equation}

{\bf 3.b) First Order Equations}

\begin{equation}\label{eq11}
 \frac{d p_1^{\mu}}{d t}=-\Gamma_{\lambda
\nu}^{\mu}\upsilon_0^{\lambda}
p_0^{\nu}-\frac{1}{2}\eta^{\mu\kappa}R_{\kappa\nu\lambda\rho}
S_0^{\lambda\rho}\upsilon_0^{\nu}+e[\eta^{\mu\kappa}F_{\kappa\beta}\upsilon_1^{\beta}+h^{\mu\kappa}F_{\kappa\beta}\upsilon_0^{\beta}]
\end{equation}

\begin{equation}\label{eq12}
\frac{dS_1^{\mu\nu}}{dt}=-\Gamma_{\lambda\rho}^{\mu}\upsilon_0^{\lambda}S_0^{\rho\nu}-\Gamma_{\lambda\rho}^{\nu}\upsilon_0^{\lambda}S_0^{\mu\rho}+[p_0^{\mu}\upsilon_1^{\nu}+
p_1^{\mu}\upsilon_0^{\nu}
-p_0^{\nu}\upsilon_1^{\mu}-p_1^{\nu}\upsilon_0^{\mu} ]
\end{equation}

\begin{eqnarray}\label{eq13}\upsilon_1^{\nu}&=&
[\frac{1}{1-eS^{\mu\nu}F_{\mu\nu}/m^2}]^{1/2}\{\frac{1}{\Delta_0^2}[2e^2N_{1a}-eN_{1b}-4e^2\frac{N_0\Delta_1}{\Delta_0}][u_0^{\nu}+\nonumber\\
&+&\frac{e}{m^2\Delta_0}F_{\kappa\sigma}S_0^{\kappa\nu}u_0^{\sigma}]\nonumber\\
&+&[1-\frac{e^2N_0}{m^4\Delta_0^2}]^{-1/2}[u_1^{\nu}-\frac{e}{2m^2\Delta_0}R_{\mu
\sigma
\lambda\rho}S_0^{\lambda\rho}S_0^{\mu\nu}u^{\sigma}\nonumber\\
&+&\frac{e}{m^2\Delta_0}F_{\kappa\sigma}(S_0^{\kappa\nu}u_1^{\sigma}+S_1^{\kappa\nu}u_0^{\sigma})
-\frac{e\Delta_1}{m^2\Delta_0^2}F_{\kappa\sigma}S_0^{\kappa\nu}u_0^{\sigma}]\}
\end{eqnarray}
where
\begin{equation}\label{eq14}
\Delta_1=\frac{1}{4m^2}R_{\sigma
\mu\lambda\rho}S_0^{\lambda\rho}S_0^{\mu\sigma
}+\frac{e}{2m^2}F_{\sigma \mu}S_1^{\sigma \mu} \end{equation}

\begin{eqnarray}\label{eq15}
N_{1a}&=&(F_{\kappa\sigma}S_0^{\kappa\nu}u_0^{\sigma})[F_{\kappa\sigma}\eta_{\nu\beta}(S_0^{\kappa\beta}u_1^{\sigma}+S_1^{\kappa\beta}u_0^{\sigma})]\nonumber\\
&+&(F_{\kappa\sigma}\eta_{\nu\beta}S_0^{\kappa\beta}u_0^{\sigma})[F_{\kappa\sigma}(S_0^{\kappa\nu}u_1^{\sigma}+S_1^{\kappa\nu}u_0^{\sigma})]
\end{eqnarray}
and
\begin{equation}\label{eq16}
N_{1b}=-2eR_{\mu\sigma\lambda\rho}[S_0^{\lambda\rho}\eta_{\nu\beta}S_0^{\mu\beta}u_0^{\sigma}(F_{\kappa\beta}S_0^{\kappa\nu}u_0^{\beta})
-S_0^{\lambda\rho}S_0^{\mu\nu}u_0^{\sigma}(F_{\kappa\sigma}\eta_{\nu\beta}S_0^{\kappa\beta}u_0^{\sigma})
\end{equation}

Because of the assumptions made in paragraph (3.a) and the form
of the metric (\ref{metric3}), the
Eqs.(\ref{eq05}),(\ref{eq14}),(\ref{eq15}) and (\ref{eq16}) give
the following results;
$N_0=N_{1a}=N_{1b}=0$,$\Delta_1=\frac{1}{2}\{h_{1,zz}[(\frac{S_0^{13}}{m})^2-(\frac{S_0^{23}}{m})^2]+2h_{1,zz}(\frac{S_0^{13}}{m})(\frac{S_0^{23}}{m})\}-\frac{eH}{m}(\frac{S_0^{12}}{m})$

\begin{equation}\label{eq17a}
\frac{d u_1^0}{dt}=0,~~\frac{d u_1^3}{d t}=0\end{equation}

\begin{equation}\label{eq17b}
\frac{d u_1^1}{d
t}+\Omega\upsilon_1^2=\frac{1}{2m}[h_{1,tz}S_0^{13}+h_{2,tz}S_0^{23}]\end{equation}

\begin{equation}\label{eq17c}
\frac{d u_1^2}{d t}-\Omega
\upsilon_1^1=-\frac{1}{2m}[h_{1,tz}S_0^{23}-h_{2,tz}S_0^{13}]\end{equation}

\begin{equation}\label{eq18a}
\frac{d S_1^{0\nu}}{d t}=m(\upsilon_1^{\nu}-u_1^{\nu}),~~\frac{d
S_1^{12}}{d t}=0\end{equation}

\begin{equation}\label{eq18b}
\frac{d S_1^{13}}{d
t}=-\frac{1}{2}[h_{1,t}S_0^{13}+h_{2,t}S_0^{23}]\end{equation} and
\begin{equation}\label{eq18c}
\frac{d S_1^{23}}{d
t}=\frac{1}{2}[h_{1,t}S_0^{23}-h_{2,t}S_0^{13}]\end{equation} The
Eqs.(\ref{pert2}) give:

\begin{equation}\label{eq19a}
\upsilon_1^0=\frac{u_1^0}{\Delta_0^{1/2}}\end{equation}

\begin{equation}\label{eq19b}
\upsilon_1^1=\frac{u_1^1}{\Delta_0^{3/2}}(1-2\Omega\frac{S_0^{12}}{m})+\frac{1}{2\Delta_0^{3/2}}[h_{1,tz}\frac{S_0^{12}}{m}\frac{S_0^{23}}{m}-h_{2,tz}\frac{S_0^{12}}{m}\frac{S_0^{13}}{m}]\end{equation}

\begin{equation}\label{eq19c}
\upsilon_1^2=\frac{u_1^2}{\Delta_0^{3/2}}(1-2\Omega\frac{S_0^{12}}{m})+\frac{1}{2\Delta_0^{3/2}}h_{1,tz}\frac{S_0^{12}}{m}[\frac{S_0^{13}}{m}-\frac{S_0^{23}}{m}]\end{equation}

\begin{eqnarray}\label{eq19d}
\upsilon_1^3&=&\frac{u_1^3}{\Delta_0^{1/2}}+\frac{1}{\Delta_0^{3/2}}\{h_{1,tz}[(\frac{S_0^{13}}{m})^2-(\frac{S_0^{23}}{m})^2]+2h_{2,tz}(\frac{S_0^{23}}{m})^2\}\nonumber\\&-&\frac{\Omega}{\Delta_0^{3/2}}[u_1^2\frac{S_0^{13}}{m}-u_1^1\frac{S_0^{23}}{m}]\end{eqnarray}

where comma means partial differentiation and
$\Omega=\frac{eH}{m}$ is the cyclotron frequency.

We solve the Eqs.(\ref{eq17a}-\ref{eq19d}) and find:

\begin{equation}\label{eq20a}
u_1^0=const.,~~u_1^3=const.\end{equation}

\begin{eqnarray}\label{eq20b}
u_1^1&=&\frac{\Delta_0^{3/2}}{D_{+}}\{[B_{10}\sin{(k_g z)}+
B_{20}\cos{(k_g z)}][\cos{(At)}-\cos{(\omega_g t)}]\nonumber\\
&+&[B_{10}\cos{(k_g z)}-B_{20}\sin{(k_g
z)}][\sin{(At)}-\sin{(\omega_g t)}]\}\end{eqnarray}

and

\begin{eqnarray}\label{eq20c}
u_1^2&=&\frac{\Delta_0^{3/2}}{D_{+}} \{[B_{10}\sin{(k_g
z)}+B_{20}\cos{(k_g z)}]\sin{(A t)}\nonumber\\&-&[B_{10}\cos{(k_g
z)}-B_{20}\sin{(k_g z)}]\cos{(At)}\}\nonumber\\
&+&\frac{\sin{(\omega_g t)}}{D_{-}}[B_{10}\sin{(k_g
z)}+B_{20}\cos{(k_g z)}]\nonumber\\&+&\frac{\cos{(\omega_g
t)}}{D_{+}}[B_{10}\cos{(k_g z)}-B_{20}\sin{(k_g z)}]
\end{eqnarray}

where

\begin{equation}\label{B10}
B_{10}=\frac{\omega_g
k_g}{2}[h_{10}(\frac{S_0^{13}}{m})+h_{20}(\frac{S_0^{23}}{m})]
-\frac{\Omega\omega_g
k_g}{2\Delta_0^{3/2}}h_{10}(\frac{S_0^{12}}{m})[\frac{S_0^{13}}{m}-\frac{S_0^{23}}{m}]
\end{equation}

\begin{equation}\label{B20}
B_{20}=-\frac{\omega_g
k_g}{2}[h_{10}(\frac{S_0^{23}}{m})-h_{20}(\frac{S_0^{13}}{m})][1-\frac{\Omega}{\Delta_0^{3/2}}(\frac{S_0^{12}}{m})]
\end{equation}

\begin{equation}\label{A}
A=\frac{\Omega}{\Delta_0^{3/2}}[1-2\Omega(\frac{S_0^{12}}{m})]\end{equation}

and
\begin{equation}\label{D}
D_{\pm}=\Omega-2\Omega^2(\frac{S_0^{12}}{m})\pm\omega_g\Delta_0^{3/2}\end{equation}

From Eqs.(\ref{eq19a}-\ref{eq19d}) and
Eqs.(\ref{eq20a}-\ref{eq20c}) we find:
$\upsilon_0^1=u_0^1=const.$ and

\begin{eqnarray}\label{solu11}
\upsilon_1^1&=&[\frac{1-2\Omega(\frac{S_0^{12})}{m}}{D_{+}}]\{[\cos{(At)}-\cos{(\omega_g
t)}][B_{10}\sin{(k_g z)}+B_{20}\cos{(k_g z)}]\nonumber\\
&+&[\sin{(At)}+\sin{(\omega_g t)}][B_{10}\cos{(k_g
z)}-B_{20}\sin{(k_g z)}]\}\nonumber\\
&+&\frac{\omega_g
k_g}{2\Delta_0^{3/2}}[h_{10}(\frac{S_0^{12}}{m})(\frac{S_0^{23}}{m})-h_{20}(\frac{S_0^{12}}{m})(\frac{S_0^{13}}{m})]e^{i(k_g
z-\omega_g t)}\end{eqnarray}

\begin{eqnarray}\label{solu12}
\upsilon_1^2&=&[\frac{1-2\Omega(\frac{S_0^{12})}{m}}{D_{+}}]\{
\sin{(At)}[B_{10}\sin{(k_g z)}+B_{20}\cos{(k_g z)}]\nonumber\\
&-&[\cos{(A t)}-\cos{(\omega_g t)}][B_{10}\cos{(k_g
z)}-B_{20}\sin{(k_g z)}]\}\nonumber\\
&+&[\frac{1-2\Omega(\frac{S_0^{12}}{m})}{D_{-}}]\sin{(\omega_g
t)}[B_{10}\sin{(k_g z)}+B_{20}\cos{(k_g z)}]\nonumber\\
&+&\frac{\omega_g
k_g}{2\Delta_0^{3/2}}(\frac{S_0^{12}}{m})[(\frac{S_0^{13}}{m})-(\frac{S_0^{23}}{m})]h_{10}e^{i(k_g
z-\omega_g t)}
\end{eqnarray}

\begin{eqnarray}\label{solu13}
\upsilon_1^3&=&\frac{u_1^3}{\Delta_0^{1/2}}-\frac{\Omega}{D_{+}}(\frac{S_0^{13}}{m})\{
\sin{(At)}[B_{10}\sin{(k_g z)}+B_{20}\cos{(k_g z)}]\nonumber\\
&-&\cos{(A t)}[B_{10}\cos{(k_g z)}-B_{20}\sin{(k_g
z)}]\}\nonumber\\
&+&\frac{\Omega}{D_{+}}(\frac{S_0^{23}}{m})\{[\cos{(At)}-\cos{(\omega_g
t)}][B_{10}\sin{(k_g z)}+B_{20}\cos{(k_g z)}]\nonumber\\
&+&[\sin{(At)}+\sin{(\omega_g t)}][B_{10}\cos{(k_g
z)}-B_{20}\sin{(k_g z)}]\}\nonumber\\
&-&\frac{\Omega}{D_{-}}(\frac{S_0^{13}}{m})\sin{(\omega_g
t)}[B_{10}\sin{(k_g z)}+B_{20}\cos{(k_g z)}]\nonumber\\
&-&\frac{\Omega}{D_{+}}(\frac{S_0^{13}}{m}) \cos{(\omega_g
t)}[B_{10}\cos{(k_g z)}-B_{20}\sin{(k_g z)}]\nonumber\\
&+&\frac{\omega
k_g}{2\Delta_0^{3/2}}\{h_{10}[(\frac{S_0^{13}}{m})^2-(\frac{S_0^{23}}{m})^2]+2h_{20}(\frac{S_0^{13}}{m})(\frac{S_0^{23}}{m})\}e^{i(k_g
z-\omega_g t)}\end{eqnarray}

{\bf Derivation of the spinning's particle trajectories:}  For
such a particle, with $\upsilon^{\mu}$ given from the above
equations, the trajectories with respect to the coordinate system
we consider and the initial conditions $t=0, X_1^{\mu}(t=0)=0$
are:

\begin{equation}\label{traj1}
X_1^{0}=\frac{u_1^0 t}{\Delta_0^{3/2}}\end{equation}

\begin{eqnarray}\label{traj2}
X_1^1&=&\frac{1-2\Omega(\frac{S_0^{12})}{m}}{D_{+}}
\{[\frac{\sin{(A t)}}{A}-\frac{\sin{(\omega_g t)}}{\omega_g
}][B_{10}\sin{(k_g z)}+B_{20}\cos{(k_g z)}]\nonumber\\
&+&[\frac{1-\cos{(\omega_g t)}}{\omega_g }+\frac{1-\cos{(A t)}}{
A}][B_{10}\cos{(k_g z)}-B_{20}\sin{(k_g z)}]\}\nonumber\\
&-&\frac{ik_g}{2\Delta_0^{3/2}}e^{ik_g z}[1-e^{-i\omega_g
t}][h_{10}(\frac{S_0^{12}}{m})(\frac{S_0^{23}}{m})-h_{20}(\frac{S_0^{12}}{m})(\frac{S_0^{13}}{m})]\end{eqnarray}

\begin{eqnarray}\label{traj3}
X_1^2&=&\frac{1-2\Omega(\frac{S_0^{12}}{m})}{D_{+}}
\{[\frac{\sin{(\omega_g t)}}{\omega_g }-\frac{\sin{(A t)}}{A
}][B_{10}\cos{(k_g z)}-B_{20}\sin{(k_g z)}]\nonumber\\
&+&[\frac{1-\cos{(At)}}{A}][B_{10}\sin{(k_g z)}+B_{20}\cos{(k_g
z)}]\}\nonumber\\
&+&[\frac{1-2\Omega(\frac{S_0^{12}}{m})}{D_{+}}][\frac{1-\cos{(\omega_g
t)}}{\omega_g }][B_{10}\sin{(k_g z)}+B_{20}\cos{(k_g
z)}]\nonumber\\
&-&\frac{ik_g}{2\Delta_0^{3/2}}e^{ik_g z}[1-e^{-i\omega_g
t}]h_{10}(\frac{S_0^{12}}{m})[(\frac{S_0^{13}}{m})-(\frac{S_0^{23}}{m})]\end{eqnarray}

and

\begin{eqnarray}\label{traj4}
X_1^3&=&\frac{u_1^3
t}{\Delta_0^{3/2}}+[\frac{\Omega}{D_{+}}(\frac{S_0^{13}}{m})
]\{[\frac{\sin{(A t)}}{A }-\frac{\sin{(\omega_g t)}}{\omega_g
}][B_{10}\cos{(k_g z)}-B_{20}\sin{(k_g z)}]\nonumber\\
&-&[\frac{1-\cos{(A t)}}{A}][B_{10}\sin{(k_g z)}+B_{20}\cos{(k_g
z)}]\}\nonumber\\
&-&[\frac{\Omega}{D_{-}}(\frac{S_0^{13}}{m})][\frac{1-\cos{(\omega_g
t)}}{\omega_g }][B_{10}\sin{(k_g z)}+B_{20}\cos{(k_g
z)}]\nonumber\\
&+&[\frac{\Omega}{D_{+}}(\frac{S_0^{23}}{m})]\{[\frac{\sin{(A
t)}}{A}-\frac{\sin{(\omega_g t)}}{\omega_g }][B_{10}\sin{(k_g
z)}+B_{20}\cos{(k_g z)}]\nonumber\\
&+&[\frac{1-\cos{(A t)}}{ A}+\frac{1-\cos{(\omega_g t)}}{
\omega_g}][B_{10}\cos{(k_g z)}-B_{20}\sin{(k_g z)}]\}\nonumber\\
&+&\frac{ik_g}{2\Delta_0^{3/2}}e^{ik_g z}[1-e^{-i\omega_g
t}]\{h_{10}[(\frac{S_0^{13}}{m})^2-(\frac{S_0^{23}}{m})^2]+2h_{20}(\frac{S_0^{13}}{m})(\frac{S_0^{23}}{m})\}\end{eqnarray}

{\bf Derivation of the spin equations:} Integrating
Eqs.(\ref{eq18a}-\ref{eq18c}) we find the components of the
$S_1^{\mu\nu}$ tensor which are:

\begin{eqnarray}\label{spin11}
\frac{S_1^{01}}{m}&=&\frac{1}{D_{+}}[1-2\Omega(\frac{S_0^{12}}{m})-\Delta_0^{3/2}]\{[\frac{\sin{(At)}}{A}-\frac{\sin{(\omega_g
t)}}{\omega_g}][B_{10}\sin{(k_g z)}+B_{20}\cos{(k_g
z)}]\nonumber\\
&+&[\frac{1-\cos{(At)}}{A}+\frac{1-\cos{(\omega_g
t)}}{\omega_g}][B_{10}\cos{(k_g z)}-B_{20}\sin{(k_g
z)}]\}\nonumber\\
&-&\frac{ik_g}{2\Delta_0^{3/2}}(\frac{S_0^{12}}{m})e^{ik_g z}
[1-e^{-i\omega_g
t}][h_{10}(\frac{S_0^{23}}{m})-h_{20}(\frac{S_0^{13}}{m})]\end{eqnarray}

\begin{eqnarray}\label{spin12}
\frac{S_1^{02}}{m}&=&\frac{1}{D_{+}}[1-2\Omega(\frac{S_0^{12}}{m})-\Delta_0^{3/2}]\{[\frac{1-\cos{(At)}}{A}][B_{10}\sin{(k_g
z)}+B_{20}\cos{(k_g z)}]\nonumber\\
&-&[\frac{\sin{(At)}}{A}][B_{10}\cos{(k_g z)}-B_{20}\sin{(k_g
z)}]\}\nonumber\\
&+&[1-2\Omega(\frac{S_0^{12}}{m})-\Delta_0^{3/2}]\{[\frac{1-\cos{(\omega_g
t)}}{\omega_g D_{-}}][B_{10}\sin{(k_g z)}+B_{20}\cos{(k_g
z)}]\nonumber\\
&+&[\frac{\sin{(\omega_gt)}}{\omega_g D_{+}}][B_{10}\cos{(k_g
z)}-B_{20}\sin{(k_g z)}]\}\nonumber\\
&-&\frac{ik_g}{2\Delta_0^{3/2}}e^{ik_g z}[1-e^{-i\omega_g
t}]h_{10}(\frac{S_0^{12}}{m})[(\frac{S_0^{13}}{m})-(\frac{S_0^{23}}{m})]\end{eqnarray}

\begin{eqnarray}\label{spin13}
\frac{S_1^{03}}{m}&=&\frac{\Omega}{D_{+}}(\frac{S_0^{23}}{m})\{[\frac{\sin{(At)}}{A}-\frac{\sin{(\omega_g
t)}}{\omega_g}][B_{10}\sin{(k_g z)}+B_{20}\cos{(k_g
z)}]\nonumber\\
&+&[\frac{1-\cos{(At)}}{A}+\frac{1-\cos{(\omega_g
t)}}{\omega_g}][B_{10}\cos{(k_g z)}-B_{20}\sin{(k_g
z)}]\}\nonumber\\
&-&\frac{\Omega}{D_{+}}(\frac{S_0^{13}}{m})\{[\frac{1-\cos{(At)}}{A}][B_{10}\sin{(k_g
z)}+B_{20}\cos{(k_g z)}]\nonumber\\
&-&[\frac{\sin{(At)}}{A}][B_{10}\cos{(k_g z)}-B_{20}\sin{(k_g
z)}]\}\nonumber\\
&-&\frac{\Omega}{D_{-}}(\frac{S_0^{13}}{m})[\frac{1-\cos{(\omega_gt)}}{\omega_g}][B_{10}\sin{(k_g
z)}+B_{20}\cos{(k_g z)}]\nonumber\\
&-&\frac{\Omega}{D_{+}}(\frac{S_0^{13}}{m})[\frac{\sin{(\omega_g
t)}}{\omega_g}][B_{10}\cos{(k_g z)}-B_{20}\sin{(k_g
z)}]\nonumber\\
&-&\frac{ik_g}{2\Delta_0^{3/2}}e^{ik_g z}[1-e^{-i\omega_g
t}]\{h_{10}[(\frac{S_0^{13}}{m})^2-(\frac{S_0^{23}}{m})^2]+2h_{20}(\frac{S_0^{13}}{m})(\frac{S_0^{23}}{m})\}\end{eqnarray}

\begin{equation}\label{spin14}
\frac{S_1^{12}}{m}=constant
\end{equation}

\begin{equation}\label{spin15}
\frac{S_1^{13}}{m}=\frac{1}{2}[h_{10}(\frac{S_0^{13}}{m})+h_{20}(\frac{S_0^{23}}{m})]e^{ik_g
z} [1-e^{-i\omega_g t}]\end{equation} and
\begin{equation}\label{spin16}
\frac{S_1^{23}}{m}=-\frac{1}{2}[h_{10}(\frac{S_0^{23}}{m})-h_{20}(\frac{S_0^{13}}{m})]e^{ik_g
z} [1-e^{-i\omega_g t}]
\end{equation}

From the Eqs.(\ref{spin11}-\ref{spin16}) we verify that because of
the GW we have non-zero first order electric and magnetic moments
of the spinning charged particle and in the absence of the GW,
all these components disappear. The electric moments particularly
exhibit resonances because in the denominators of the
Eqs.(\ref{spin11}-\ref{spin13}) appear the expressions
$\Delta_0=1-\frac{eH}{m}\frac{S_0^{12}}{m}$,
$A=\frac{\Omega}{\Delta_0^{3/2}}[1-2\Omega(\frac{S_0^{12}}{m})]$
and
$D_{\pm}=\Omega-2\Omega^2(\frac{S_0^{12}}{m})\pm\omega_g\Delta_0^{3/2}$,
which become zero for certain values of the Larmor frequency
$\Omega=\frac{eH}{m}$, the ratio $\frac{S_0^{12}}{m}$ and angular
frequency of the GW, $\omega_g $. We notify that while the
electric moments of the spinning charged particle exhibit such an
interesting behavior, the magnetic moments are independent from
the magnetic field and the $S_0^{12}$ zero order component of the
spin tensor. Also, for the same reasons mentioned above the
Eqs.(\ref{solu11}-\ref{solu13}) exhibit resonaces. In the
neighborhood of those resonaces the charged spinning particle
gains energy from the GW and accelerates radiating. The above
comments become more plausible by examining a special case of the
Eqs.(\ref{spin11}-\ref{spin16}) and
Eqs.(\ref{solu11}-\ref{solu13}) in the appendix .

\newpage
\section*{4. Discussion}

Dealing with the interaction of a GW with a spinning particle in
the presence of a uniform magnetic field in the linearized theory
of general relativity, we found the following results:

1) In the case where the GW and magnetic field are across the z
axis, the components of the 4-velocity, 4-momentum and the spin
tensor $S^{\mu\nu}$, exhibit resonance at
$\Omega=(\frac{S_0^{12}}{m})^{-1}$ $(\Delta_0=0)$. Due to the
co-existence of the constant magnetic field with GW, a strong
coupling between the frequency $\Omega=\frac{eH}{m}$ and the
magnetic moment $S_0^{12}$ of the charged spinning particle
occur. This coupling gives rise to the above resonance.

Also, for the same reasons mentioned above, some other resonances
appear in the
Eqs.(\ref{eq20b}-\ref{eq20c}),Eqs.(\ref{B10}-\ref{A}),
Eqs.(\ref{solu11}-\ref{solu13}),Eqs.(\ref{traj1}-\ref{traj4}) and
Eqs.(\ref{spin11}-\ref{spin16}) which are solutions to 4-order
polynomial in terms of $\Omega$; $D_{\pm}=0\Rightarrow
4\Omega^4(\frac{S_0^{12}}{m})^2-4\Omega^3[\omega_g^2(\frac{S_0^{12}}{m})^3
-4\frac{S_0^{12}}{m})]+\Omega^2[1-3\omega_g^2(\frac{S_0^{12}}{m})^2]
+3\omega_g^2(\frac{S_0^{12}}{m})\Omega-\omega_g^2=0$.

2) It is interesting to notify that in the absence of the GW, the
magnetic field and the components of the spin tensor decouple and
the magnetic resonances disappear. In this case, where the GW
does not exist(see ref.33 and references there in), the motion of
a spinning charged point-particle of mass $m$ and charge $q$ is
described in an 4-dimensional Minkowski space time by its position
$X^{\mu}(t)$, defining the particle's world line, its 4-velocity
$u^{\mu}$, which is tangent to the world-line and its
polarization tensor $D_{\mu\nu}(t)$, an antisymmetric 4-tensor
which combines the intrinsic magnetic dipole moment \textbf{M}( a
pseudo 3-vector) with the intrinsic electric dipole moment
\textbf{d} (a real 3-vector) at every given point of the
world-line through the relations $
D_{ij}=\frac{1}{c}\epsilon_{ijk}M_k$ and $-iD_{i4}=d_i$ where
$(i,j,k)=1,2,3$. In the absence of external fields, the intrinsic
dipole moments are found from the values of \textbf{M} and
\textbf{d} (rest frame of the free particle). Usually we are
interested in charged particle with no intrinsic electric dipole
moment in the rest frame of the free particle. This may expressed
by the condition $D_{\mu\nu} u^{\nu}=0$. On the other hand, the
polarization tensor is relate to an intrinsic angular momentum
tensor $S_{\mu\nu}$(spin tensor) through the expression
$D_{\mu\nu}=(q/mc)S_{\mu\nu}$. From the above mentioned equations
we have the condition $S_{\mu\nu}u^{\nu}=0$. When this relation
holds, $S_{\mu\nu}$ is space-like with only 3 non-zero components
in the rest frame of the free particle e.g.
$S_{ij}^{(0)}=\epsilon_{ijk}s^{k}$ and $S_{i0}^{(0)}=0$. Besides,
we have to point out that in the case of the unperturbed Minkowski
space-time the classical spin is introduced somehow indirectly,
via the electromagnetic polarization tensor, because the empirical
meaning of classical magnetic and electric dipole moments is
clear.

3) In the case that the GW does exist and in the limit of the high
frequency approximation[37,38], the charged particle which initial
is at rest, starts to have a combination of an orbital and
spinning motion, described by the Eqs.(\ref{traj1}-\ref{traj4})
and Eqs.(\ref{spin11}-\ref{spin16}), respectively. The spin
tensor $S^{\mu\nu}=S_0^{\mu\nu}+S_1^{\mu\nu}$, exhibits electric
and magnetic moments eventhough initially had magnetic moments
only. Because of the Eqs.(\ref{eq18a}), the electric moments
exhibit the same resonances as the components of the 4-momentum
and in the neighborhood of those resonaces energy is transferred
from the GW to the spinning particle. The magnetic moments do not
depend neither on the magnetic field nor on the component
$S^{12}$. Under these circumstances one could hope to detect such
GW.

A possible astrophysical environment where the interaction
studied in this paper maybe be of relevance is the binary neutron
star merger. In this scenario, two magnetized neutron stars
merge, forming (if the equation of state allows it) a very
massive, differentially rotating object and a possible low-mass
disk around it (the object could survive for hundreds of seconds
before collapsing to a black hole [39]. The magnetosphere of this
object will be rotating rapidly and be filled with plasma, while
near the object, gravitational waves of large amplitude will be
emitted. It would be interesting to study the conditions under
which the interaction studied in the present paper could lead to
observable phenomena during such a binary neutron star merger.

To make some further comments related to possible astrophysical
application of the Eqs.(51-65), we have to consider the
Pauli-Lubanski covariant spin vector formula $
S_\sigma=\frac{1}{2}\epsilon_{\rho\mu\nu\sigma}
u^{\rho}S^{\mu\nu} $, which gives
$S_0^{12}=S_0^3,~~S_0^{13}=-S_0^2,~~S_0^{23}=S_0^1 $ and assume,
for simplicity, that e.g. $S_0^3=S_0^1=S$, while $S_0^2=0$, then
some typical values for this scenario of some astrophysical
importance may be met when, for example, the amplitude of the GW
is $h\approx 10^{-10}$, $H\approx 10^6 G$, and for an electron
one has roughly $S\approx 10^{-13}$m [1]. Such conditions can be
found around various compact objects, for example near neutron
stars which posses a magnetic field of the order $10^8-10^{12} G$
and emit GW due to glitches or rotational instabilities excited
by accretion (see Ref.[40,41,42])

{\bf Acknowledgements:} The author would like to express his
gratitude to Enric Verdaguer for his comments, criticism and
interesting references. Also the author would like to thank
Loukas Vlahos, Kostas Kokkotas and Nick Stergioulas, for their
helpful suggestions and discussions on this topic.

\newpage

\section*{Appendix}

In the appendix we will present a special case of the Eqs.(53-55)
and Eqs.(60-67).

We assume $z=0$ and for further simplicity we obtain
$h_{10}=h_{20}=h_0$ and $S_0^{13}=0$. Because of the assumption
$c=G=1$, $k_g=\omega_g$. Now Eqs.(53-52) read:

\begin{eqnarray}
\upsilon_1^1&=&\frac{\omega_g^2}{2D_{+}}h_0(\frac{S_0^{23}}{m})[1-2\Omega(\frac{S_0^{12}}{m})]\{[\cos{(\omega_g
t)}-\cos{(At)}][1-\frac{\Omega}{\Delta_0^{2/3}}(\frac{S_0^{12}}{m})]\nonumber\\
&+&[\sin{(\omega_g
t)}+\sin{(At)}][1+\frac{\Omega}{\Delta_0^{2/3}}(\frac{S_0^{12}}{m})]\}\nonumber\\
&+&h_0\sin{(\omega_g
t)}\frac{\omega_g^2}{2\Delta_0^{2/3}}(\frac{S_0^{23}}{m})(\frac{S_0^{12}}{m})\end{eqnarray}

\begin{eqnarray}
\upsilon_1^2&=&-\frac{\omega_g^2}{2D_{+}}h_0(\frac{S_0^{23}}{m})[1-2\Omega(\frac{S_0^{12}}{m})]\{\sin{(At)}][1-\frac{\Omega}{\Delta_0^{2/3}}(\frac{S_0^{12}}{m})]\nonumber\\
&+&[\cos{(At)}-\cos{(\omega_g t)}][1+\frac{\Omega}{\Delta_0^{2/3}}(\frac{S_0^{12}}{m})]\}\nonumber\\
&-&h_0\sin{(\omega_gt)}\frac{\omega_g^2}{2D_{-}}(\frac{S_0^{23}}{m})[1-2\Omega(\frac{S_0^{12}}{m})][1-\frac{\Omega}{\Delta_0^{2/3}}(\frac{S_0^{12}}{m})]\nonumber\\
&-&h_0\cos{(\omega_g
t)}\frac{\omega_g^2}{2\Delta_0^{2/3}}(\frac{S_0^{23}}{m})(\frac{S_0^{12}}{m})\end{eqnarray}

\begin{eqnarray}
\upsilon_1^3&=& \frac{u_1^3}{\Delta_0^{1/2}}+\frac{\Omega\omega_g}{2D_{+}}h_0(\frac{S_0^{23}}{m})^2\{[\cos{(\omega_gt)}-\cos{(At)}][1-\frac{\Omega}{\Delta_0^{2/3}}(\frac{S_0^{12}}{m})]\nonumber\\
&+&[\sin{(\omega_gt)}+\sin{(At)}][1+\frac{\Omega}{\Delta_0^{2/3}}(\frac{S_0^{12}}{m})]\}\nonumber\\
&-&h_0\cos{(\omega_g
t)}\frac{\omega_g^2}{2\Delta_0^{2/3}}(\frac{S_0^{23}}{m})^2\end{eqnarray}

Also the Eqs.(60-67) become:

\begin{eqnarray}
\frac{S_1^{01}}{m}&=&\frac{\omega_g^2}{2D_{+}}h_0(\frac{S_0^{23}}{m})[1-2\Omega(\frac{S_0^{12}}{m})-\Delta_0^{3/2}]\{[\frac{\sin{(\omega_g
t)}}{\omega_g}-\frac{\sin{(At)}}{A}][1-\frac{\Omega}{\Delta_0^{2/3}}(\frac{S_0^{12}}{m})]\nonumber\\
&+&[\frac{1-\cos{(\omega_g
t)}}{\omega_g}+\frac{1-\cos{(At)}}{A}][1+\frac{\Omega}{\Delta_0^{2/3}}(\frac{S_0^{12}}{m})]\}\nonumber\\
&-&h_0\sin{(\omega_g
t)}\frac{\omega_g}{2\Delta_0^{2/3}}(\frac{S_0^{23}}{m})(\frac{S_0^{12}}{m})\end{eqnarray}

\begin{eqnarray}
\frac{S_1^{02}}{m}&=&-\frac{\omega_g^2}{2D_{+}}h_0(\frac{S_0^{23}}{m})[1-2\Omega(\frac{S_0^{12}}{m})-\Delta_0^{3/2}]\{\frac{1-\cos{(At)}}{A}[1-\frac{\Omega}{\Delta_0^{2/3}}(\frac{S_0^{12}}{m})]\nonumber\\
&+&\frac{\sin{(At)}}{A}[1+\frac{\Omega}{\Delta_0^{2/3}}(\frac{S_0^{12}}{m})]\}\nonumber\\
&-&\frac{\omega_g^2}{2}h_0(\frac{S_0^{23}}{m})[1-2\Omega(\frac{S_0^{12}}{m})-\Delta_0^{3/2}]\{\frac{1-\cos{(\omega_g t)}}{\omega_gD_{-}}[1-\frac{\Omega}{\Delta_0^{2/3}}(\frac{S_0^{12}}{m})]\nonumber\\
&-&\frac{\sin{(\omega_gt)}}{\omega_g D{+}}[1+\frac{\Omega}{\Delta_0^{2/3}}(\frac{S_0^{12}}{m})]\}\nonumber\\
&-&h_0\sin{(\omega_g
t)}\frac{\omega_g}{2\Delta_0^{2/3}}(\frac{S_0^{23}}{m})(\frac{S_0^{12}}{m})\end{eqnarray}

\begin{eqnarray}
\frac{S_1^{03}}{m}&=&\frac{\Omega\omega_g^2}{2D_{+}}h_0(\frac{S_0^{23}}{m})^2\{[\frac{\sin{(\omega_g
t)}}{\omega_g}-\frac{\sin{(At)}}{A}][1-\frac{\Omega}{\Delta_0^{2/3}}(\frac{S_0^{12}}{m})]\nonumber\\
&+&[\frac{1-\cos{(\omega_g
t)}}{\omega_g}+\frac{1-\cos{(At)}}{A}][1+\frac{\Omega}{\Delta_0^{2/3}}(\frac{S_0^{12}}{m})]\}\nonumber\\
&-&h_0\sin{(\omega_g
t)}\frac{\omega_g}{2\Delta_0^{2/3}}(\frac{S_0^{23}}{m})^2\end{eqnarray}

\begin{equation}
\frac{S_1^{12}}{m}=const.\end{equation}

\begin{equation}
\frac{S_1^{13}}{m}=-\frac{S_1^{23}}{m}=\frac{1}{2}h_0(\frac{S_0^{23}}{m})[1-\cos{(\omega_g
t)}]\end{equation}

\end{document}